# On the Role of Minor Galaxy Mergers in the Formation of Active Galactic Nuclei

Michael R. Corbin

Steward Observatory

The University of Arizona

Tucson, AZ, 85721; mcorbin@as.arizona.edu

ABSTRACT

The large-scale (~ 100 kpc) environments of Seyfert galaxies are not significantly different from those of non-Seyfert galaxies. In the context of the interaction model of the formation of active galactic nuclei (AGN), it has thus been proposed that AGN form via "minor mergers" of large disk galaxies with smaller companions. We test this hypothesis by comparing the nuclear spectra of 105 bright nearby galaxies with measurements of their $R$ or $r$ band morphological asymmetries at three successive radii. We find no significant differences between these asymmetries among the 13 Seyfert galaxies in the sample and galaxies having other nuclear spectral types (absorption, H II region - like, LINER), nor is there strong qualitative evidence that such mergers have occurred among any of the Seyferts or LINERs. Thus either any minor mergers began $> 1$ Gyr ago and are essentially complete, or they did not occur at all, and AGN form independently of any type of interaction. Support for the latter interpretation is provided by the growing evidence that supermassive black holes exist in the cores of most elliptical and early-type spiral galaxies, which in turn suggests that nuclear activity represents a normal phase in the evolution of the bulges of massive galaxies. Galaxy mergers may increase the luminosity of Seyfert nuclei to the level of QSOs, which could explain why the latter objects appear to be found in rich environments and in interacting systems.

1. INTRODUCTION

It has long been suspected that galaxy interactions play a role in the formation and/or fueling of active galactic nuclei (AGN). Early studies, including Petrosian (1982), Dahari (1984; see also Dahari 1985a, 1985b) and Keel et al. (1985) noted that Seyfert galaxies appear to show an excess of close companion galaxies relative to their non-active counterparts. However, more recent studies using larger samples better matched in redshift, morphology and luminosity have led to an erosion of support for the interaction model of AGN. Specifically, comparitive analyses of the environments of Seyfert and non-Seyfert galaxies by Laurikainen & Salo (1995), De Robertis, Hayhoe & Yee (1998), De Robertis, Yee & Hayhoe (1998) and Dultzin-Hacyan et al. (1999) find no statistically significant differences between the ~ 100 kpc - scale environments of Seyferts and non-Seyferts, although, interestingly, Seyfert 2 galaxies (those lacking broad emission lines) appear to inhabit richer environments than Seyfert 1 galaxies (those displaying both broad and narrow lines). De Robertis, Yee & Hayhoe (1998) have thus argued that if interactions trigger AGN, they most likely do so through "minor mergers" of the host galaxies with smaller companions. A *prima facie* case for this model can be made on the basis of the evidence that approximately 75% of isolated spiral galaxies appear to have at least one close satellite galaxy (Zaritsky et al. 1997), and that the incidence of spiral galaxies in which a minor merger appears to have recently (< 1 Gyr ) occurred is ~ 20% - 30% (Zaritsky & Rix 1997; Rudnick & Rix 1998), which is comparable to the incidence of Seyfert activity in the general galaxy population (e.g. Ho, Filippenko & Sargent 1995, 1997a). Such mergers also appear to boost the star formation rate of the larger galaxy (Rudnick, Rix & Kennicutt 2000) which could be related to the formation and/or fueling of a nuclear black hole.

A true test of the minor merger model requires a quantitative comparison of the morphologies of a large, well-defined sample of galaxies with their nuclear spectral types. The means for such a comparison have recently been provided by Conselice et al. (2000; hereafter C2000), who investigate the morphological asymmetries of 113 bright, nearby galaxies using the deep CCD images of Frei et al.

(1996). These asymmetry measurements appear to be a useful extension of traditional classification systems, and are a sensitive indicator of dynamically disturbed systems. C2000 note the nuclear spectral types of some of the galaxies in their sample, including several Seyferts and galaxies with low-ionization nuclear emission-line regions (LINERs). Optical spectra of most of the remaining galaxies are available in the spectroscopic atlas of 486 bright Northern galaxies presented by Ho et al. (1995), and the spectra of other objects are available in the literature. In this *Letter* we combine these data for a statistical comparison of the C2000 asymmetry measurements with the galaxy nuclear spectral types. The main goal is to see if the 13 Seyfert galaxies in this sample display larger asymmetries than galaxies having other nuclear spectral types, which would be indicative of recent minor mergers. *N*-body simulations of minor mergers by Walker, Mihos & Hernquist (1996) show that they produce marked disturbances in the morphologies of the larger galaxy, at least within ~ 1 Gyr of the onset of the merger, and the asymmetry measurements of C2000 should be sensitive to these differences. This is a preliminary study to a more extensive comparison of galaxy morphological properties and nuclear spectral type based on deep *BVR* imaging of most of the 486 galaxies in the Ho et al. (1995) sample that is currently in progress.

## 2. DATA

C2000 use the $B_J$, $R$, and Thuan-Gunn $g$, $r$ and $i$ band CCD images of Frei et al. (1996). The Frei et al. sample galaxies are resolved down to ~ 100 pc and cover the full range of Hubble types, but are not necessarily representative of the entire local galaxy population. For the present analysis, we use the asymmetry measurements of C2000 made from a 180° rotation of the $R$ and $r$ band images of the galaxies,

$$A = \frac{\sum |(I_0 - I_{180})|}{2\sum |I_0|}$$

where a correction for noise in the image background has also been applied. The $R$ and $r$ band images were selected to better trace the older stellar populations within the galaxies. Following Petrosian (1976), C2000 measure this parameter at three successive radii, defined in fractions of the galaxy mean intensity, $\eta( r ) \equiv I( r ) / < I( r ) >$, with $\eta = 1$ at the galaxy center and approaching zero at large radii. C2000 also measure the asymmetry at three successive radii defined by the curve of growth of the galaxy intensity profile.

To classify the galaxy nuclear spectra, we use the notes provided in Tables 1 and 2 of C2000, as well as the atlas of Ho et al. (1995) and sources in the literature. We divide the spectra into four types: absorption, H II region-like, Seyfert and LINERs. We note that in many objects the spectra are more complicated, e.g. composite H II region-like + LINER systems, but for simplicity and because of the relatively small sample size we chose a single type depending on which spectral features are dominant. The sample contains both Seyfert 1 and Seyfert 2 galaxies, but the small total number of Seyferts (13) does not allow separate statistical tests of the two types. We are able to assign nuclear spectral types to 105 of the 113 galaxies in the C2000 sample. In Figure 1 we show contour plots of the Frei et al. (1996) *R* or *r* band CCD images of four of the galaxies in the sample, one of each nuclear spectral type.

3. RESULTS

We applied the two-sided Kolmogorov-Smirnov (K-S) test to the C2000 asymmetry measurements for each sub-sample of galaxies in comparison with the remaining galaxies. We also performed this test on the T parameter values (the numerical index of morphological type) for the galaxies as listed in the Third Reference Catalog of Bright Galaxies (de Vaucouleurs et al. 1991). The mean values and K-S test results are summarized in Table 1. The results for the asymmetry measurements based on the curve-of-growth parameterization of the galaxy intensities were found to be approximately the same as those based on $\eta$, and so are not listed. The mean values follow the correlation of the T parameter and asymmetry values found by C2000. Indeed, the differences in the asymmetry values for each galaxy nuclear spectral type revealed by the K-S test must be interpreted in the context of these morphological differences, e.g., galaxies with absorption spectra are almost exclusively elliptical or S0 galaxies, which tend to be dynamically relaxed with very symmetric structure (e.g. NGC 4621, Fig. 1). Similarly, galaxies with H II

-like spectra tend to be late-type spirals, which are often asymmetric in their outer regions and have a flocculent appearance due to dust within the galaxy disk (e.g. NGC 3893, Fig. 1). In this respect it is possible to say most generally that galaxy nuclear spectral type and asymmetry are more strongly dependent on morphological type than any other parameter.

While the Seyfert sample is relatively small, there is no indication of a higher level of asymmetry at any radii among them than in the other spectral classes, or of any strong deviation from the T-type / asymmetry correlation. Indeed, the most asymmetric galaxies in the sample are interacting systems which have H II - like spectra (see C2000). In Figure 2 we show the distributions of the asymmetry measurement at the innermost radius, $A[\eta(0.8)]$, for each of the sub-samples, where it can be seen that the distribution of the Seyferts matches well that of the other galaxies, particularly the LINERs and non-interacting H II - like galaxies. We additionally inspected the Frei et al. (1996) images of the Seyferts and LINERs in the sample for qualitative evidence of recent minor mergers, but found no clear cases of such (e.g., note the basic symmetry of the Seyfert 2 galaxy NGC 3486, Fig. 1).

## 4. DISCUSSION

The lack of evidence for strong asymmetries in the Seyfert galaxies of the sample in comparison with the numerical simulations of Walker et al. (1996) indicates that either any minor mergers began > 1 Gyr ago and are now essentially complete, or else that they have not occurred at all, and the Seyfert nuclei formed independently of such mergers. In conjunction with the lack of environmental differences between Seyferts and non-Seyferts noted in § 1, if the latter interpretation is correct, then the interaction model of AGN formation is now seriously called into question. It does not seem possible to conclusively rule out the former interpretation, given that there are no strong observational constraints on the lifetimes of Seyfert nuclei. However, if minor mergers trigger AGN, then it is possible to say that they appear to do so only in the late stages of the mergers, when the remnants of the smaller galaxy have been absorbed into the bulge of the larger galaxy. This would then imply that the nuclei have been active only within the last ~ 0.1 Gyr (see Walker et al. 1996).

As noted in § 2, it now appears that galaxy nuclear spectra, as well as morphological asymmetry, are more strongly related to morphological type than environment. Consistent with this result is the fact that Seyfert galaxies are predominantly early-type spirals (Moles, Márquez & Pérez 1995; Ho, Filippenko & Sargent 1997a). Recent studies of the stellar kinematics and radio emission of the cores of nearby elliptical and S0 galaxies, many of which are included in the present sample, have also indicated that nearly all contain dormant or weakly accreting supermassive black holes (Magorrian et al. 1998; Ho 1999). Similar results have been obtained for galaxies in the Local Group, including the Milky Way (e.g. Genzel et al. 1997), M 31 (e.g. Kormendy & Bender 1999) and M 32 (e.g. van der Marel et al. 1998). The nature of the LINER phenomenon, whether a manifestation of AGN or stellar activity or both, remains controversial. However, since at least part of the LINER population does indeed appear to represent the low end of the AGN luminosity function (e.g. Ho et al. 1997c; Colina & Koratkar 1997), the high frequency of LINERs among the local galaxy population (see Table 1; Ho et al. 1997a) offers further evidence that a large fraction of spiral galaxy cores contain massive or supermassive black holes. Therefore, unless minor mergers also occur in all such galaxies, one is led to a picture similar to that investigated by Norman & Scoville (1988) in which the formation of a massive or supermassive black hole occurs as part of the natural evolution of most or all massive galaxy cores, without requiring galaxy interaction. We could thus be witnessing an evolutionary sequence in which ellipticals and S0 galaxies have formed their holes and finished their active phase, early-type spirals are close to or in their active phase and can be classified as Seyferts or LINERs, and late-type spirals showing H II - like nuclei have yet to form, or will not form, holes sufficiently massive for significant non-thermal activity. In support of such an evolutionary model is the evidence that the Hubble sequence, to which nuclear activity is related, is also a sequence of star-formation rate (Kennicutt 1992), with late-type spirals being most actively star forming. Similarly, Veilleux et al. (1995) find that the H$\beta$ and Mg I absorption features in the spectra of luminous infrared galaxies hosting AGN are stronger than those in luminous infrared galaxies only showing evidence of current star formation, suggesting that the stellar populations of the former galaxies are older. Also in support of the starburst evolution model is the evidence that the fuel source of AGN is local, i.e. not driven in from kiloparsec scales by tidal interactions with other galaxies (see the references in § 1), nor by bar instabilities ( McLeod & Rieke 1995; Ho, Filippenko & Sargent 1997b; Regan & Mulchaey 1999).

Even if interactions and mergers are not required to form AGN, they may still play an important role in their fueling. The luminosities of Seyfert nuclei could be driven up to the levels of QSOs by the deposition of additional fuel onto existing black holes, thereby increasing their accretion rates. This would account for the evidence that QSOs and quasars inhabit richer environments than Seyfert galaxies (e.g. Ellingson, Yee & Green 1991; Hutchings 1995; Hutchings, Crampton & Johnson 1995), and are often found in strongly interacting systems (e.g. Hines et al. 1999). QSO evolution could then naturally occur due to the cosmological evolution of the galaxy merger rate (see Carlberg 1990).

I thank Chris Conselice for discussion of his asymmetry measurements, Luis Ho and Greg Rudnick for helpful discussions of the topics addressed in this paper, and an anonymous referee for comments that improved it. This work was supported by NASA grant NAG 5-3042 to The University of Arizona.

TABLE 1

MEAN T-TYPE AND *R, r* BAND ASYMMETRY VALUES AND KOLMOGOROV-SMIRNOV

TEST RESULTS

|  | T | $A[\eta(0.8)]$ | $A[\eta(0.5)]$ | $A[\eta(0.2)]$ |
|---|---|---|---|---|
| Sample & No. of Objects | | Mean Values | | |
| 1. Absorption (12) | -2.6 | 0.030 | 0.030 | 0.039 |
| 2. H II - like (48) | 5.0 | 0.080 | 0.146 | 0.200 |
| 3. Seyfert (13) | 3.2 | 0.060 | 0.128 | 0.147 |
| 4. LINER (32) | 0.8 | 0.040 | 0.064 | 0.081 |
| Comparison | | K-S Test Results[1] | | |
| 1 vs. 2 + 3 + 4 | >99.99 | 99.33 | >99.99 | >99.99 |
| 2 vs. 1 + 3 + 4 | >99.99 | 99.88 | >99.99 | >99.99 |
| 3 vs. 1 + 2 + 4 | 76.30 | 82.24 | 70.34 | 79.82 |
| 4 vs. 1 + 2 + 3 | >99.99 | 97.45 | 99.95 | 99.91 |

[1] Probability that the sample values are drawn from different parent distributions.

FIGURE CAPTIONS

FIG. 1. -- Contour plots of four galaxies in the present sample, with the nuclear spectral type indicated. Images are oriented with north at the top and east to the left. The images are taken from Frei et al. (1996) and are in the *R* band for NGC 4621, NGC 3893 and NGC 3486, and the Gunn *r* band for NGC 4548. Contours are spaced 0.25 magnitudes apart, and were begun approximately 10% above the sky level.

FIG. 2. – Comparison of the *R* and *r* band asymmetry measurements of the galaxies in the present sample at the $\eta = 0.8$ radii. A similar overlap in distributions between galaxies of different nuclear spectral types occurs at the $\eta = 0.5$ and $\eta = 0.2$ radii (see Table 1).